\documentclass[aps,prl,showpacs,twocolumn,groupedaddress]{revtex4-1}

\pdfoutput=1
\usepackage{amsmath}
\usepackage{cancel}
\usepackage{float}
\usepackage{ifsym}
\usepackage{color}
\usepackage{latexsym}
\usepackage{latexsym}
\usepackage{dcolumn}
\usepackage{epsfig}
\usepackage{amsfonts}
\usepackage{amsmath}
\usepackage{dsfont}
\usepackage{accents}
\usepackage{longtable}

\usepackage[all,cmtip]{xy}

\usepackage{bm}
\usepackage{graphicx}

\usepackage{hyperref} 

\newcommand{\be}{{\underline{e}}}

\newcommand{\beps}{{\underline{\eps}}}

\newcommand{\bX}{{\bm{X}}}
\newcommand{\bY}{{\bm{Y}}}

\newcommand{\ba}{{\underline{a}}}
\newcommand{\bal}{{\underline{\alpha}}}

\newcommand{\bQ}{{\bm{Q}}}

\newcommand{\Id}{{\mathds{1}}}
\newcommand{\al}{\alpha}

\newcommand{\la}{\lambda}
\newcommand{\eps}{\epsilon}

\newcommand{\etab}{{\underline{\eta}}}

\newcommand{\wh}{\widehat}

\newcommand{\dwh}[2]{{\skew{#1}\widehat{ \widehat #2}}}

\begin{document}

\title{Application of Finite Strain Landau Theory To High Pressure Phase Transitions}
\author{A.~Tr\"oster}
\affiliation{Vienna University of Technology, Wiedner Hauptstrasse 8-10/136, A-1040 Wien, Austria}
\email{andreas.troester@tuwien.ac.at}
\author{W.~Schranz}
\affiliation{University of Vienna, Boltzmanngasse 5, A-1090 Vienna, Austria}
\author{ F.~Karsai and P.~Blaha}
\affiliation{Vienna University of Technology, Getreidemarkt 9/165-TC, A-1060 Vienna, Austria}
\date{\today}

\begin{abstract}
In this paper we explain how to set up what is in fact the only possible consistent construction scheme for a Landau theory of high pressure 
phase transitions that systematically allows to take into account elastic nonlinearities. We also show how to incorporate available information
on the pressure dependence of elastic constants taken from experiment or simulation.
We apply our new theory to the example of the high pressure cubic-tetragonal phase transition in Strontium Titanate,
a model perovskite that has played a central role in the development of the theory of structural phase transitions. 
Armed with pressure dependent elastic constants calculated by density functional theory, we give a both qualitatively as well as quantitatively 
satisfying description of recent high precision experimental data. Our nonlinear theory also allows to predict a number of additional elastic transition anomalies 
that are accessible to experiment.
\end{abstract}


\maketitle
   
High pressure phase transitions in crystals constitute a central research area of modern physics that continues to
attract widespread interest ranging from astrophysics and geology to chemistry and nanotechnology.   
Experimentally, this is due to the permanent refinement of diamond anvil cell techniques,
while the ongoing theoretical and hardware-related advances also allow quite precise  \emph{ab initio}
calculations of high pressure transitions. In comparison, even nowadays the sophistication of the theoretical concepts
which are employed to analyze and interpret the high quality data produced by these methods still leaves a lot to be desired.

The reason for this deplorable situation is not hard to see. In condensed matter physics,
the group-theoretic analysis of symmetry changes at phase transitions is a central concept. 
Combined with thermodynamics, the resulting machinery of irreducible representations, order parameters (OPs), domain patterns etc., which comes under the name of Landau theory (LT), 
has proven its value countless times as one of the most useful and versatile approaches to gain 
both a qualitative as well as a quantitative understanding of phase transitions.
In the LT of solid state structural phase transitions, \emph{strain} usually plays an important role as a primary or secondary OP \cite{ToledanoToledano_LTPT_1987}.
For temperature-driven structural transitions and/or at small applied external pressures, 
strain effects may be small enough to allow the involved strain components to be treated as infinitesimal.
The elastic energy may then be truncated beyond harmonic order. Consistency then demands to sacrifice the possibility of a pressure dependence of the ``bare'' elastic constants
characterizing the high symmetry phase. 
However, once the external pressure comes close to the value of the crystal's elastic constants (typically some $100$\,GPa), 
ignoring this pressure dependence and other nonlinear effects is bound to result in errors that can quickly approach $100\%$. 

Despite these obvious shortcomings, infinitesimal strains continue to be frequently 
employed in the Landau analysis of high pressure transition data \cite{Carpenter_JGR105_2000,Guennou_JPCM23_485901_2011,Bouvier_JPCM14_3981_2002,Errandonea_EPL77_56001_2007,Togo_PRB_78_134106_2008}. 
Trying to circumvent the technical and conceptual difficulties 
of nonlinear elasticity theory, many authors still resort to mathematical sledgehammer methods. 
Of course, bold manipulations like fitting volume data to a nonlinear equation of state 
while continuing to treat strains as infinitesimal and introducing pressure dependent elastic constants in an ad hoc manner
inevitably yield inconsistencies which must then be swept under the rug. For instance, the
tensorial consistency relation between the pressure-dependence of the elastic constants and that of the unit cell parameters \cite{KoppensteinerTroester_PRB74_2006}
are generously ignored in all brute force attempts to introduce an ad hoc $P$-dependence to the ``bare'' elastic constants $C^0_{ij}$. 
Even more important, nonlinear elasticity theory carefully distinguishes between second derivatives of the free energy taken with respect to various strain measures,
and the resulting ``elastic constants'', which are known as e.g.~Birch, Voigt and Huang coefficients \cite{Wallace_SSP25_1970} differ from each other by terms of order $P$. 
Erroneous use of these may thus easily introduce dramatic errors into the analysis of high pressure data (cf.~the discussion \cite{MuserSchoffel_PRL_90_079701_2003}). 
Such misconceptions are particularly disastrous to eigenvalue calculations like in applications of the Born stability criteria (for a recent example see Ref.~\cite{Togo_PRB_78_134106_2008}). 
Inconsistencies of the naive approach even arise on the basic conceptual level
since both strain and stress appear as parameters in the resulting formalism, whereas one should be the control and the other the dependent variable. 
It is the purpose of the present paper to show how to consistently solve all of these problems and to present the general construction of a LT coupled to 
finite strain. 

Progress in this direction has already been made more than ten years ago.
In Ref.~\cite{TroesterSchranzMiletich_PRL88_2002} the concept of LT coupled to finite strain has been developed and successfully applied to 
describe experimental data. 
The central idea of this approach was that even if the total observed strain at a high pressure phase transition may appear to be 
far from being ``infinitesimal'' and must therefore be described in terms of a proper nonlinear strain measure such as the Lagrangian strain tensor
$\etab$, its actual \emph{spontaneous} contribution $\underline{\wh\eps}$ originating from the emergence of a nonzero equilibrium value $\bar{\bQ}=\bar{\bQ}(P)$ 
of the OP $\bQ$ is still bound to be ``small'' near a second order or weakly first order phase transition.
To separate $\underline{\wh\eps}$ from $\etab$, three reference systems 
connected by the scheme
$
\xymatrix{
\bX  \ar[r]^{\ba,\be} \ar@/_0.8pc/[rr]_{\bal,\etab}&\wh\bX \ar[r]^{\wh\bal,\wh\beps} &\dwh{3}{\bX} 
}
$ were introduced in Ref.~\cite{TroesterSchranzMiletich_PRL88_2002}:
(i) the fully deformed system $\dwh{3}{\bX}$ (denoted as $\wh\bY$ in Ref.~\cite{TroesterSchranzMiletich_PRL88_2002}) at pressure $P$ and relaxed equilibrium value $\bar{\bQ}=\bar{\bQ}(P)$.
(ii) the undeformed zero pressure ``laboratory/ambient system/frame'' denoted by $\bX$, relative to which the state $\dwh{3}{\bX}$ corresponds to
the experimentally measured deformation tensor $\bal$ and Lagrangian strain $\etab=(1/2)(\bal^T\cdot\bal-\Id)$. 
(iii) a ``background reference system'' $\wh\bX$ defined as the (hypothetical) state of the system at pressure $P$ and
clamped OP $\bQ\equiv0$. Relative to $\wh\bX$, one would thus precisely measure the spontaneous strain $\wh\beps=(1/2)(\wh\bal^T\cdot\wh\bal-\Id)$ accompanying a deformation gradient matrix $\wh\bal$.
$\wh\bX$ and $\bX$ are related through a deformation gradient tensor with components $a_{ij}=\partial\wh X_i/\partial X_j$
and the resulting Lagrangian strain tensor with components $e_{ij}=(1/2)(a_{ki} a_{kj}-\delta_{ij})$ as $\be=(1/2)(\ba^T\cdot\ba-\Id)$. 
Given these definitions, the total experimentally observed strain is thus decomposed
into the nonlinear \cite{Wallace_TDC_1998} superposition $\eta_{ij}=e_{ij}+a_{ki}\wh\eps_{kl}a_{lj}$.
In the background reference frame $\wh\bX$, the $\bQ$-independent elastic contribution $F_0(\wh\beps;\wh\bX)$ to the Landau free energy was assumed to be captured by the harmonic approximation 
\begin{eqnarray}
\frac{F_0(\wh\beps;\wh\bX)}{V[\wh\bX]}\approx
\sum_{ij}\tau_{ij}\wh\eps_{ij}
+\frac{1}{2}\sum_{ijkl}C_{ijkl}[\wh\bX]\wh\eps_{ij}\wh\eps_{kl}
\label{eqn:xnjnqwnqwnqwjqwjqwq}
\end{eqnarray}
involving the external stress tensor $\tau_{ij}$ and the elastic constants $C_{ijkl}[\wh\bX]$ of the background system.
Furthermore, in accordance with the traditional approach of LT, the fact that both OP as well as strain components remain small in the vicinity of the transition suggested
to drop all coupling terms between strain and OP beyond second order in $\bQ$. For simplicity, we take the OP to be scalar
and content ourselves with a single coupling between $Q$ and $\wh{\beps}$ of type $Q^2d_{ij}(\wh\bX)\wh\eps_{ij}$, which yields the
ansatz
\begin{eqnarray}
\frac{F(Q,\wh\beps;\wh\bX)}{V(\wh\bX)}
=\Phi(Q;\wh\bX)+Q^2d_{ij}(\wh\bX)\wh\eps_{ij}+\frac{F_0(\wh\beps;\wh\bX)}{V[\wh\bX]}
\label{eqn:xxmmqmklqmkmkqmqkmqk}
\end{eqnarray}
in which we introduced the potential density $\Phi(Q;\wh\bX)$  of the pure OP contribution.
In what follows we shall assume $\Phi(Q;\wh\bX)\equiv\frac{A[\wh\bX]}{2}Q^2+\frac{B[\wh\bX]}{4}Q^4+\frac{C[\wh\bX]}{6}Q^6$ to be a simple sixth order polynomial in $Q$.
Unfortunately, even with these assumptions the resulting strain equilibrium conditions
\begin{eqnarray}
\sum_{mn}\frac{\bar{\wh\al}_{mi}\bar{\wh\al}_{nj}}{J[\bar{\wh\bal}]}\left(
\bar Q^2 d_{mn}(\wh\bX)+C_{mnkl}[\wh\bX]\wh\eps_{kl} \right)\approx 0
\label{eqn:GGLandauinbXmitAnsatzcjjnejwdnqejd}
\end{eqnarray}
assumed the highly nonlinear form \cite{TroesterSchranzMiletich_PRL88_2002}
where $J(\wh\bal)=V[\dwh{3}{\bX}]/V[\wh\bX]=\det\wh\bal$.
To overcome this problem, in Ref.~\cite{TroesterSchranzMiletich_PRL88_2002} the spontaneous strain $\wh{\beps}^2$ was assumed to be infinitesimal, 
and the nonlinear prefactors $\frac{\bar{\wh\al}_{mi}\bar{\wh\al}_{nj}}{J[\bar{\wh\bal}]}$ were consequently put to one. One then arrives at a system of 
linear equations that must be solved for the strain components $\wh\eps_{kl}$ as functions of $\bar Q$ by inverting the tensor $C_{mnkl}[\wh\bX]$. 
However, this step is delicate, because the required invertibility is not automatically guaranteed. Indeed, we recently realized that one can do a lot better.

In fact, let us continue to regard $\bar{\wh\beps}$ as a full Lagrangian strain tensor, even though its components may be numerically small. 
Using the first order approximations \cite{MorrisKrenn2000}
$\bar{\wh\al}_{ij}=\delta_{ij}+\bar{\wh\eps}_{ij}+O(\bar{\wh\eps}^2)$ and $J[\bar{\wh\bal}]= 1+\sum_{k}\bar{\wh\eps}_{kk}+O(\bar{\wh\eps}^2)$
we expand the geometrical prefactor in (\ref{eqn:GGLandauinbXmitAnsatzcjjnejwdnqejd}) 
up to harmonic order in $\bar{\wh\beps}$. A short calculation results in
\begin{eqnarray}
\bar Q^2 d_{ij}[\wh\bX]+ \sum_{kl}B_{ijkl}[\wh\bX]\bar{\wh\eps}_{kl}\equiv&0
\label{eqn:GGLandauinbXmitAnsatz1}
\end{eqnarray}
in which the well-known 
\emph{Birch coefficients} 
$B_{ijkl}[\wh\bX]=C_{ijkl}[\wh\bX]
+\frac{1}{2}\left(\tau_{jk}\delta_{il}+\tau_{ik}\delta_{jl}+\tau_{jl}\delta_{ik}+\tau_{il}\delta_{jk}-2\tau_{ij}\delta_{kl}\right)$
of the background system $\wh\bX$ have taken over the role formerly played by the elastic constants~\cite{Wallace_TDC_1998,MorrisKrenn2000}.
Since the background system $\wh\bX$ is defined by the constraint $\bar Q\equiv0$, which inhibits the transition
under investigation, application of the Born stability criteria \cite{Born_JCP_7_591_1939,WallacePR_162_776_1967,Wang_Yip_PRB_52_12627_1995,ZhouJoos_PRB_54_3841_1996,MorrisKrenn2000,WangJPCM_24_245402_2012}
now ensures that its tensorial inverse
\begin{eqnarray*}
\sum_{mn}B_{ijmn}[\wh\bX]S_{mnkl}[\wh\bX]=\frac{1}{2}(\delta_{ik}\delta_{jl}+\delta_{il}\delta_{jk}),   
\label{eqn:xmskxmqkxmkxmkqmxkmxkqmxkmxkqmqkmx}
\end{eqnarray*}
the tensor $S_{ijkl}[\wh\bX]$ of elastic compliances, exists. 
Inserting the solution of (\ref{eqn:GGLandauinbXmitAnsatz1})
\begin{eqnarray}
\bar{\wh\eps}_{mn}=-\bar Q^2 \sum_{ij}d_{ij}(\wh\bX)S_{mnij}[\wh\bX]
\label{eqn:loesungwheps}
\end{eqnarray}
into the second equilibrium condition 
\begin{eqnarray}
\Phi'(\bar Q;\wh\bX)+2\bar Q\sum_{ij}d_{ij}(\wh\bX)\bar{\wh\eps}_{ij}&\equiv& 0
\label{eqn:GGLandauinbXmitAnsatz2}
\end{eqnarray}
and re-integrating by $\bar Q$ finally yields (up to an unimportant constant) the \emph{renormalized background pure OP potential density}
\begin{eqnarray}
\Phi_{R}(Q;\wh\bX)&=&\frac{A_R[\wh\bX]}{2}Q^2+\frac{B_R[\wh\bX]}{4}Q^4+\frac{C_R[\wh\bX]}{6}Q^6
\end{eqnarray}
where $A_R[\wh\bX]=A[\wh\bX]$ and $C_R[\wh\bX]=C[\wh\bX]$ but
\begin{eqnarray}
B_R[\wh\bX]=B[\wh\bX]-2\sum_{ijkl}d_{ij}[\wh\bX]S_{ijkl}[\wh\bX]d_{kl}[\wh\bX]\label{eqn:xcnbcqbbchbhbhbchbcb} 
\end{eqnarray}
The equilibrium condition for the OP then takes the simple form
$\Phi_{R}'(\bar Q;\wh\bX)\equiv 0$. All of these quantities are defined with respect to the reference state $\wh\bX=\wh\bX[P]$.
It remains to determine this implicit $P$-dependence. As to the elastic constants $C_{ijkl}[\wh\bX]$,
this information is in principle accessible by density functional theory (DFT) calculations. The key observation that allows to assess the remaining implicit $P$-dependence of 
the coefficients $A[\wh\bX],B[\wh\bX],C[\wh\bX]$ and $d_{ij}[\wh\bX]$ is the following. Working in the laboratory system $\bX$, 
one would be forced to go to prohibitively high powers in the Landau expansion 
\begin{widetext}
\begin{eqnarray}
F(Q,\etab;\bX)&\equiv&
V(\bX)\Bigg[
\frac{A}{2}Q^2+\frac{B}{4}Q^4+\frac{C}{6}Q^2+\dots
+\frac{1}{2!}\sum_{ijkl}C^{(2)}_{ijkl}\eta_{ij}\eta_{kl}+\frac{1}{3!}\sum_{ijklmn}C^{(3)}_{ijklmn}\eta_{ij}\eta_{kl}\eta_{mn}+\dots
\\&&\phantom{V(\bX)\Bigg[}
+\sum_{N=1}^{\infty}Q^{2N}\left(\sum_{ij}d^{(2N,1)}_{ij}\eta_{ij}+\frac{1}{2!}\sum_{ijkl}d^{(2N,2)}_{ijkl}\eta_{ij}\eta_{kl}+\frac{1}{3!}\sum_{ijklmn}d^{(2N,3)}_{ijklmn}\eta_{ij}\eta_{kl}\eta_{mn}+\dots\right)\nonumber
\Bigg]
\label{eqn:xmklsxmsxmkqxmkxmkxmkxkxxmqxmxqkx}
\end{eqnarray}
\end{widetext}
to capture nonlinear elastic effects with sufficient precision. Nevertheless, by definition all its coefficients 
are \emph{strain-independent} (but possibly $T$-dependent) \emph{constants}.
We now compare common coefficients for the monomials $Q^{2N}\wh\eps_{i_1j_1} \dots\wh\eps_{i_nj_n}$ in a combined
expansion of the invariance relation 
\begin{eqnarray}
F(Q,\wh{\beps};\wh{\bX})\quad\equiv\quad F(Q,\be+\ba^{+}\cdot\wh\beps\cdot\ba;\bX)  
\label{eqn:inKomponentenX}
\end{eqnarray}
of the free energy under a change of the strain reference frame. Guided by the requirement that the maximum power of $Q$ appearing on both sides should be identical (i.e.~six in our present model),
a lengthy calculation, whose details will be given elsewhere, results in 
\begin{eqnarray}
J(\ba)d_{st}[\wh\bX]&=&\sum_{ij}a_{si}d^{(2,1)}_{ij}a_{tj}+\sum_{ijkl}a_{si}d^{(2,2)}_{ijkl}a_{tj}e_{kl}+\dots
\nonumber 
\\
J(\ba)A(\wh\bX)&=&A+2\sum_{ij}d^{(2,1)}_{ij}e_{ij}+\sum_{ijkl}d^{(2,2)}_{ijkl}e_{ij}e_{kl}+\dots
\nonumber
\\J(\ba)B(\wh\bX)&=&B,\quad J(\ba)C(\hat\bX)=C.
\label{eqn:cndjkcndjcjljcndjcndjkcnsdjkclsmwmkwmwk}
\end{eqnarray}
To judge the usability of these power series in the background strains $e_{kl}$, assume that the numerical values for the coefficients 
$A,B,C,d^{(2,1)}$ and $C_{ijkl}^{(2)}$ defining the zero pressure theory are known, as is the case in many applications.
Once the background elastic constants $C^0_{ijkl}(P)\equiv C_{ijkl}[\bX(P)]$ have been measured or determined from DFT, we can also compute the components $e_{ij}$ 
(see Ref.~\cite{TroesterSchranzMiletich_PRL88_2002}). This leaves the higher order coefficients $d^{(2,\al)}_{ijkl}$ with $\al=2,3,\dots$ as free parameters in fits of experimental data.
Superficially, we may thus still face a large number of unknowns. However, the symmetry of the low pressure phase can be used to further reduce their number.
In fact, assume for simplicity that the low pressure phase is cubic. Then $e_{ij}\equiv e\delta_{ij}$ is diagonal, and the equations 
(\ref{eqn:cndjkcndjcjljcndjcndjkcnsdjkclsmwmkwmwk}) collapse to
\begin{eqnarray}
a\, d_{ij}[\wh\bX]&=&d^{(2,1)}_{ij}+d^{(2,2)}_{ij}e+d^{(2,3)}_{ij}e^2+\dots   
\label{eqn:jqjndjqdnejdnjdnejdnqejdnqejdcdcdckdcdnqedbgvdqdvq}
\\
a^3\, A(\wh\bX)&=&A+2d^{(2,1)}_{ii}e+d^{(2,2)}_{ii}e^2+\frac{d^{(2,3)}_{ii}}{3}e^3+\dots
\label{eqn:xnsjnxnjnjsnjnjnjnjnhbvgqvgcfinal}
\end{eqnarray}
in which only certain \emph{sums} 
\begin{eqnarray}
d^{(2,2)}_{ij}:=\sum_{k}d^{(2,2)}_{ijkk},\quad
d^{(2,3)}_{ij}:=\sum_{kl}d^{(2,3)}_{ijkklll},\dots
\label{eqn:sxkwmmomwomoms}
\end{eqnarray}
of these unknown coefficients remain as parameters. 

We illustrate the advantages of our present approach by 
performing a fit of recent high-precision measurements \cite{Guennou_PRB_054115_2010} of the 
pressure-induced $\mathrm{Pm\bar3m \leftrightarrow I4/mcm}$ transition around $P_c=9.6$\,GPa
in the model perovskite Strontium Titanate (STO) at room temperature. At ambient pressure, $\mathrm{Sr Ti O_3}$
was already shown in the 1960s to undergo a cubic $\to$ tetragonal
transition at $T_c\approx 105$ K recognized to be an archetypal model for other soft mode-driven
structural phase transitions \cite{Cowley_Phil_Soc_354_2799_1996}.
The crystal class of perovskites itself is also of widespread interest for a number of technological applications as non volatile computer memories,
detectors of magnetic signals, electrolytes in solid oxide fuel cells, read heads in hard disks, etc \cite{Tejuca1992}. 
Moreover, recent research \cite{Murakami_Nature_485_90_2012} indicates that more than $93\%$ by volume of the 
Earth's lower mantle consist of minerals of the perovskite structure.
\begin{figure}[tb]
\includegraphics[angle=-90,width=7.7cm]{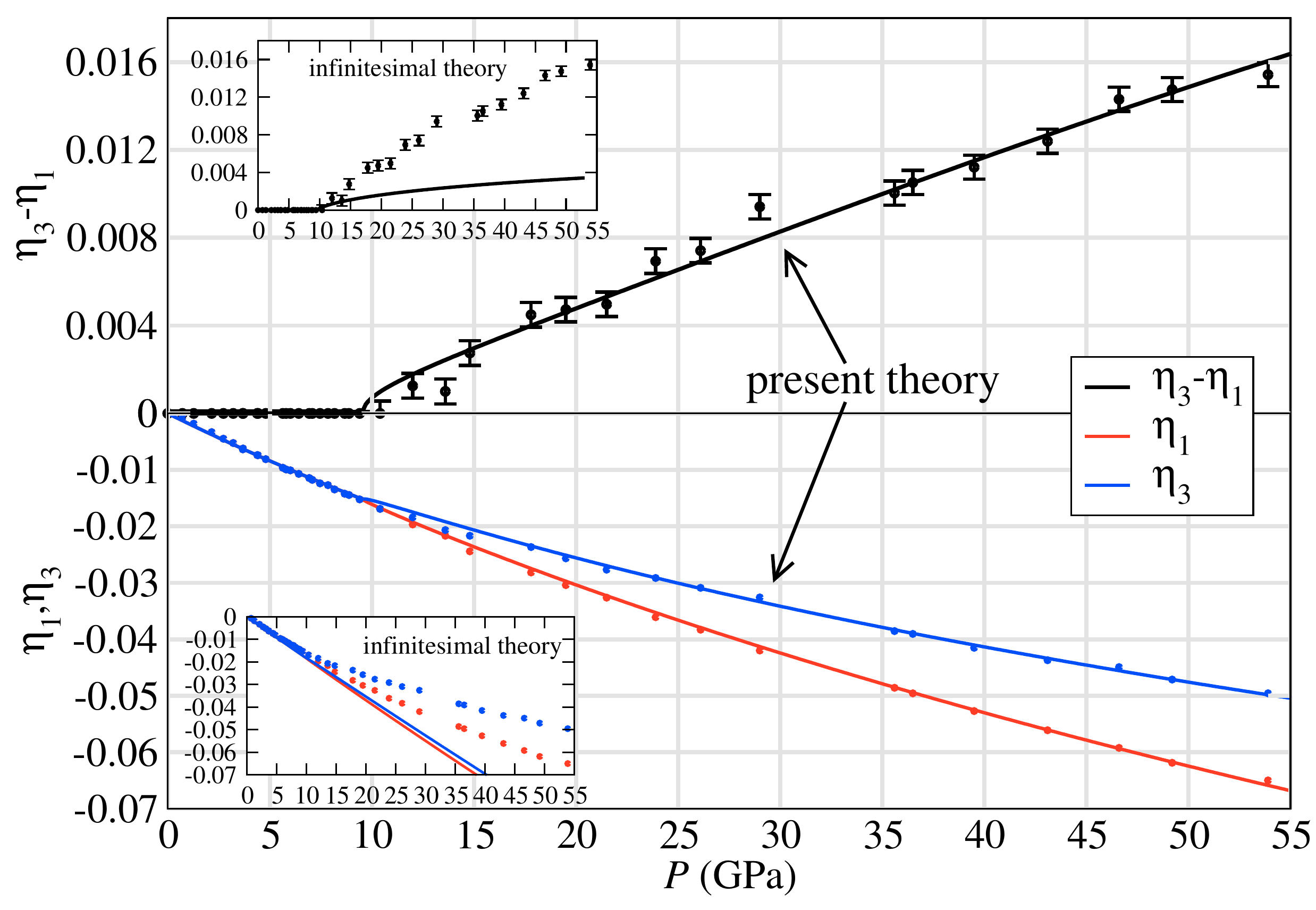}
\caption{$P$-dependence of Lagrangian strain components $\eta_1(P),\eta_3(P)$
(solid colored lines) as obtained from a fit of our theory to the experimental data of Ref.~\onlinecite{Guennou_PRB_054115_2010} (black data points). 
Insets: corresponding results of the infinitesimal approach using the parameters of Table III, left column of Ref.~\onlinecite{Guennou_PRB_054115_2010}.
\label{fig:etafit}}
 \end{figure} 

\begin{figure}[tb]
\includegraphics[angle=-90,width=7.7cm]{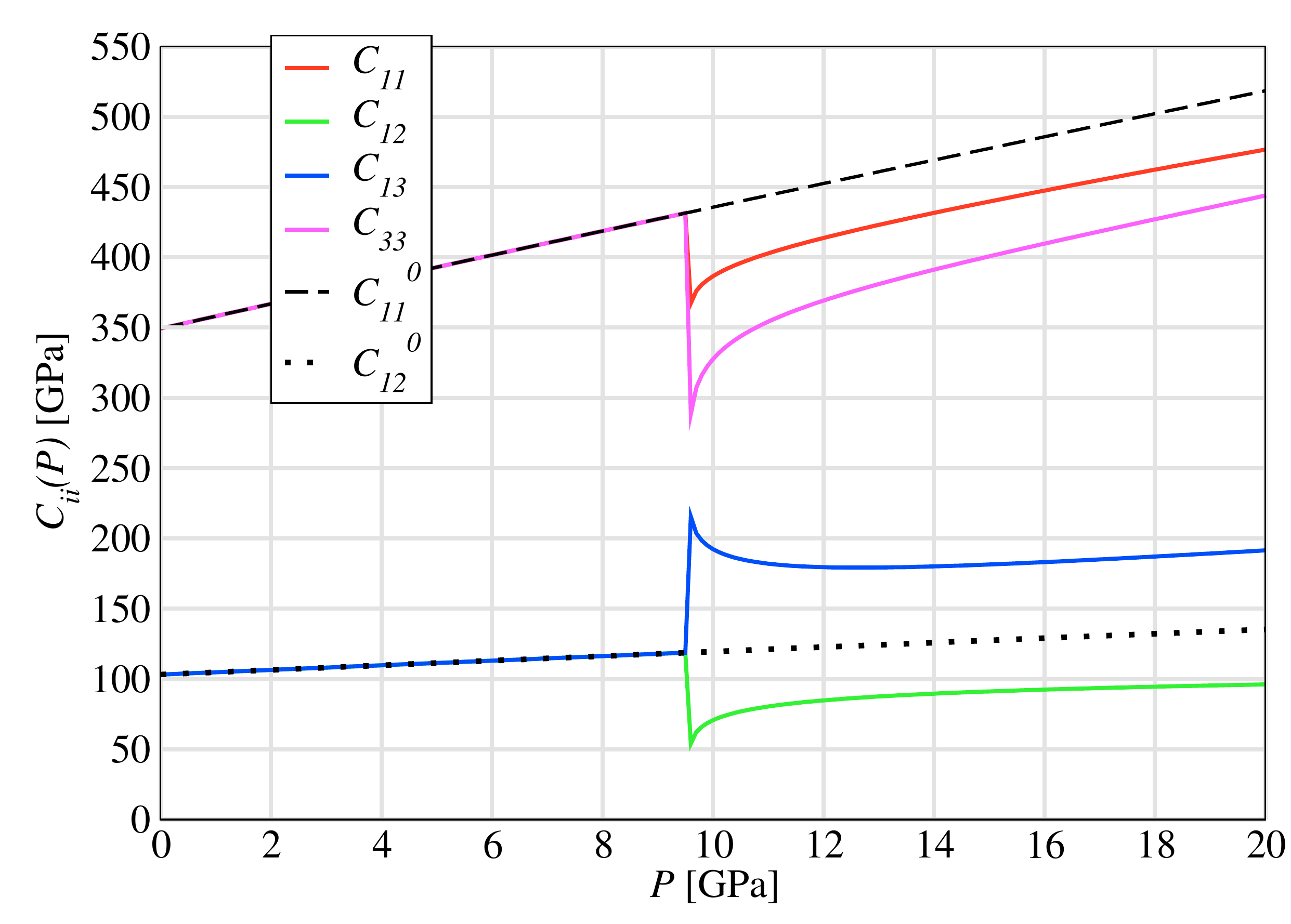}
\caption{Solid colored lines: pressure dependencies of elastic constants $C_{11}(P),\, C_{12}(P)\, C_{13}(P)$ and $C_{33}(P)$ 
as determined from combining our present theory with the approach of Ref.~\cite{SlonczewskiThomas1970}.
Black dotted lines: background elastic constants.
}
\label{fig:ciiall}
\end{figure}

A glance at Eqn.~(1) of Ref.~\cite{Guennou_PRB_054115_2010} indicates that the Landau potential underlying the transition is just of the structure discussed above,
and we immediately inherit the whole set of ambient parameters $A_0,B,C,d_{11}^{(2,1)}$ and $d_{33}^{(2,1)}$ \cite{Guennouparameters} from the left column of Table III
of Ref.~\cite{Guennou_PRB_054115_2010}. 
To provide the only missing further input for applying our present nonlinear theory, we calculate the background elastic constants $C^0_{11}(P),C^0_{12}(P)$ 
(reverting to Voigt notation) from total energies in DFT,
using the Wien2k package~\cite{WIEN2k} in combination with the PBE\_sol~\cite{Perdew_PRL100_136406_2008} functional, 
which gave the best results with respect to the experimental lattice constants.
With increasing $P$, $C_{11}(P)$ is found to increase roughly
linear from $349.3$\,GPa to about $895.9$\,GPa, while $C_{12}(P)$ increases from $103.1$\,GPa to $208.1$\,GPa between $0$ and $70\,\mathrm{GPa}$ (see the dotted lines in Fig.~\ref{fig:ciiall}). 
To the alert reader, these pronounced $P$-dependencies should suffice to cast severe doubts on any physical predictions deduced from the simple infinitesimal strain approach.
More details on the DFT calculations will be presented elsewhere.

\begin{figure}[tb]
\includegraphics[angle=-90,width=7.7cm]{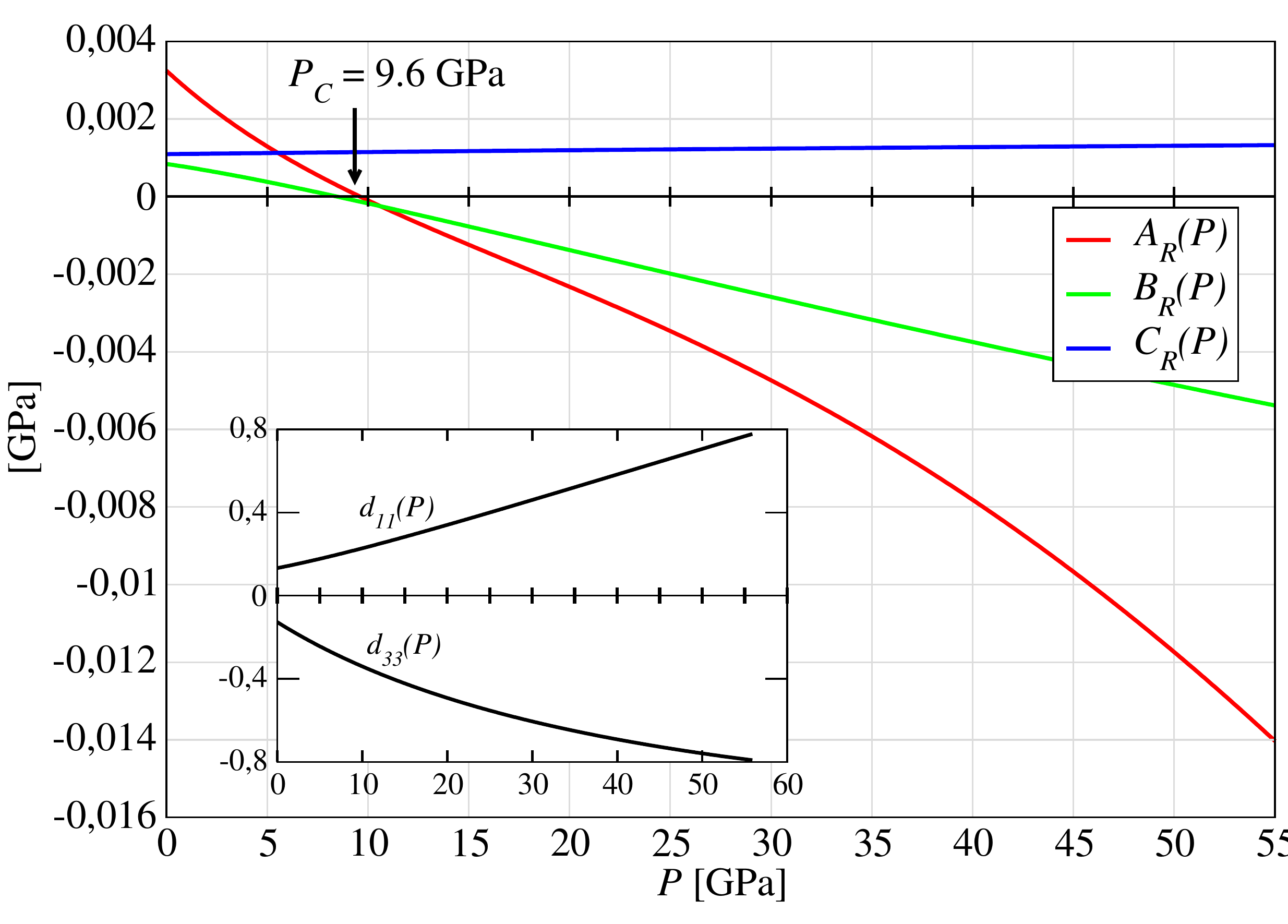}
\caption{Effective $P$-dependence of Landau coefficients $A_R(P)\equiv A_R[\wh\bX],\,B_R(P)\equiv B_R[\wh\bX],\,C_R(P)\equiv C_R[\wh\bX]$.
Inset: $P$-dependence of parameters $d_{ii}(P)\equiv d_{ii}[\wh\bX]$.}
\label{fig:abcr}
\end{figure} 
With all other quantities fixed, the remaining unknown $P$-dependence of $d_{ii}(P)\equiv d_{ii}[\wh\bX(P)]$ is accounted for by introducing four fit parameters
$d_{11}^{(2,2)},d_{33}^{(2,2)},d_{11}^{(2,3)},d_{33}^{(2,3)}$. Fixing the critical pressure to $P_c=9.6$\,GPa, the resulting implicit equation $A_R[\wh\bX(P_c)]\equiv0$  
allows to eliminate one of these. As Fig.~\ref{fig:etafit} reveals, a fit based of the remaining three parameters works extremely well.
The figure insets also exposes the complete failure of a brute force infinitesimal approach \cite{Guennou_PRB_054115_2010}. 
The $P$-dependence of $A_R(P)\equiv A_R[\bX(P)]$ obtained from the fit indeed roughly resembles the linear slope one would expect from standard LT (see Fig.~\ref{fig:abcr}). However,
due to the strong $P$-dependence exhibited by the bare elastic constants, $B_R(P)\equiv B_R[\bX(P)]$ also decreases strongly with increasing $P$ and crosses 
zero near $P_c$. From an infinitesimal approach, we could have neither deduced the tricritical or weakly first order character of the transition implied by this behavior of $B_R(P)$,
nor could we have anticipated the considerable but smooth growth in modulus of the couplings $d_{11}(P)$ and $d_{33}(P)$ over the considered pressure range shown in the inset of Fig.~\ref{fig:abcr}. 

In summary, our present nonlinear LT offers a complete quantitative description of the experimentally measured strains and their transition anomalies.
In addition it also allows to study the $P$-dependence 
of OP, soft mode frequency, superlattice scattering intensities and a number of other experimentally accessible observables. 
While a full discussion of these calculations must be deferred to a longer publication, here we content ourselves with reporting our prediction of 
the transition anomalies exhibited by the set of static longitudinal elastic constants of STO in Fig.~\ref{fig:ciiall}, which were obtained from generalizing the approach of 
Ref.~\cite{SlonczewskiThomas1970} to the nonlinear regime.

A.T.~and W.S.~acknowledge support by the Austrian Science Fund (FWF) Projects P22087-N16 and P32982-N20, respectively. F.K.~and P.B.~acknowledge the TU Vienna doctoral college COMPMAT.

\end{document}